\documentclass[twocolumn,showpacs,preprintnumbers,amsmath,%
amssymb,prl,superscriptaddress]{revtex4}

\usepackage{graphicx}
\usepackage{color}

\usepackage[pdfsubject={electronic structure},
pdfauthor={H. Gretarsson},pdftitle={X-ray induced electronic structure change in CuIr$_2$S$_4$},%
pdfpagetransition=Glitter]{hyperref}

\def\HG#1 {\emph{\color{blue}#1}}
\begin{document}

\title{X-ray induced electronic structure change in CuIr$_2$S$_4$}

\author{H.~Gretarsson}
\affiliation{Department of Physics, University of Toronto, 60
St.~George St., Toronto, ON, Canada M5S 1A7}
\author{Jungho~Kim}
\affiliation{CMC-XOR, Advanced Photon Source, Argonne National Laboratory, Argonne, Illinois 60439, USA}
\author{D.~Casa}
\affiliation{CMC-XOR, Advanced Photon Source, Argonne National Laboratory, Argonne, Illinois 60439, USA}
\author{T.~Gog}
\affiliation{CMC-XOR, Advanced Photon Source, Argonne National
Laboratory, Argonne, Illinois 60439, USA}
\author{K.~R.~Choi}
\affiliation{l\_PEM, Pohang University of Science and Technology,
Pohang 790-784, Korea}
\author{S.~W.~Cheong}
\affiliation{l\_PEM, Pohang University of Science and Technology,
Pohang 790-784, Korea} \affiliation{R-CEM and Department of Physics
and Astronomy, Rutgers University, Piscataway, New Jersey 08854,
USA}
\author{Young-June~Kim}
\email{yjkim@physics.utoronto.ca}
\affiliation{Department of
Physics, University of Toronto, 60 St.~George St., Toronto, ON,
Canada M5S 1A7}

\date{\today}

\begin{abstract}
The electronic structure of CuIr$_2$S$_4$ has been investigated using
various bulk-sensitive x-ray spectroscopic methods near the Ir
$L_3$-edge: resonant inelastic x-ray scattering (RIXS), x-ray
absorption spectroscopy in the partial fluorescence yield (PFY-XAS)
mode, and resonant x-ray emission spectroscopy (RXES). A strong RIXS
signal (0.75~eV) resulting from a charge-density-wave gap opening is
observed below the metal-insulator transition temperature of 230~K.
The resultant modification of electronic structure is consistent
with the density functional theory prediction. In the spin- and
charge- dimer disordered phase induced by x-ray irradiation below
50~K, we find that a broad peak around 0.4~eV appears in the RIXS
spectrum.
\end{abstract}


\maketitle

The role of relativistic spin-orbit interactions in condensed matter
systems is drawing much attention recently, and there are intensive
research efforts to elucidate physics of materials with 5d electrons
\cite{Okamoto2007,BJKIM2008,Pesin2010,Shitade2009,Jackeli2009}. A
thiospinel CuIr$_2$S$_4$ is such an example. The high temperature
metallic phase has a cubic spinel structure \cite{Furubayashi1994},
but the system becomes insulating and diamagnetic with triclinic
structure below the metal-insulator transition (MIT) temperature of
T$_{\rm{MI}} \sim 226$~K \cite{Radaelli2002}. The ``A"-site in the
spinel structure is occupied by Cu, which is monovalent
\cite{Matsuno1997,Croft2003,Kijima2009}, and therefore inactive near
the Fermi energy. The mixed-valent iridium ions (Ir$^{3.5+}$) occupy
the B-sites, and form a pyrochlore lattice. From their powder neutron
diffraction experiments, Radaelli and coworkers proposed an
intriguing octamer model of charge ordering, in which Ir$^{4+}$ ions
form spin- and structural- dimers separated by non-magnetic
Ir$^{3+}$ ions \cite{Radaelli2002}. This unusual charge-order (CO)
was explained by Khomskii and Mizokawa as an orbitally driven
Peierls transition. That is, a large overlap of Ir d orbitals along
the (110)-type direction renders the band structure
quasi-one-dimensional and susceptible to Peierls instability
\cite{Khomskii2005}.

What is mystifying is its behavior when the sample is irradiated
with x-rays or electrons. A significant drop in resistivity was
observed when the sample was exposed to x-ray or electron
irradiation at low temperatures (T $<$ 50~K)
\cite{Ishibashi2002,Furubayashi2003}. In
addition, the resistivity change occurs quite slowly, over a period of several
minutes, and this effect could be reversed by heating the sample
above 100~K. Detailed single crystal x-ray scattering studies showed
that the long-range CO was destroyed and a short-range
incommensurate CO appeared in this irradiation-induced
low-temperature (IILT) phase \cite{Kiryukhin2006}. Although
spectroscopic data suggest that substantial electronic structure
modification occurs at the MIT, no spectroscopic changes in this
IILT phase below 50 K have been reported to date \cite{Takubo2005,Takubo2008}. For example,
opening of an insulating gap of 0.15~eV below the MIT temperature
was observed in both optics \cite{Wang2004} and photoemission experiments
\cite{Takubo2005}, but the photoemission spectra did not
show any change in the IILT phase \cite{Takubo2005,Takubo2008}.

X-ray spectroscopy provides a natural avenue to investigate this
intriguing behavior, since x-rays can act as a radiation source as
well as a probing particle. However, detailed investigation of
electronic structure using core level x-ray spectroscopy has been
limited. Although several Ir 4f photoemission experiments have
reported that Ir$^{3+}$/Ir$^{4+}$ charge disproportion does exist in
the insulating phase, the high surface sensitivity of this technique obfuscates the interpretation \cite{Takubo2005,Takubo2008,Noh2007}. While x-ray absorption
spectroscopy (XAS) at the S $K$-edge has been valuable in
detecting redistribution of S 3p states across the MIT
\cite{Croft2003}, the Ir $L_3$-edge XAS has not been very useful due to
broad spectral features arising from the short core-hole lifetime
\cite{Croft2003,Kijima2009}. In this work, we report
bulk-sensitive x-ray spectroscopic studies which allow us to obtain direct
information regarding the Ir 5d electronic states across the MIT and
also in the IILT phase. Utilizing resonant inelastic x-ray
scattering (RIXS) and resonant x-ray emission spectroscopy (RXES),
we observe that the Ir 5d \emph{t$_{2g}$} band shifts as a result of the
opening of an insulating gap in the CDW phase, which confirms density
functional theory (DFT) calculations
\cite{Oda1995,Sasaki2004,Sarkar2009}. We also find that a
mid-infrared peak in the RIXS spectrum emerges around 0.4~eV in the
IILT phase, suggesting that not only the crystal \cite{Kiryukhin2006} but also the electronic structure are modified due to the x-ray irradiation. To explain this experimental observation, we propose a model in which local dimers freed from the long-range CO can hop around.

\begin{figure}[htb]
\includegraphics[width=\columnwidth]{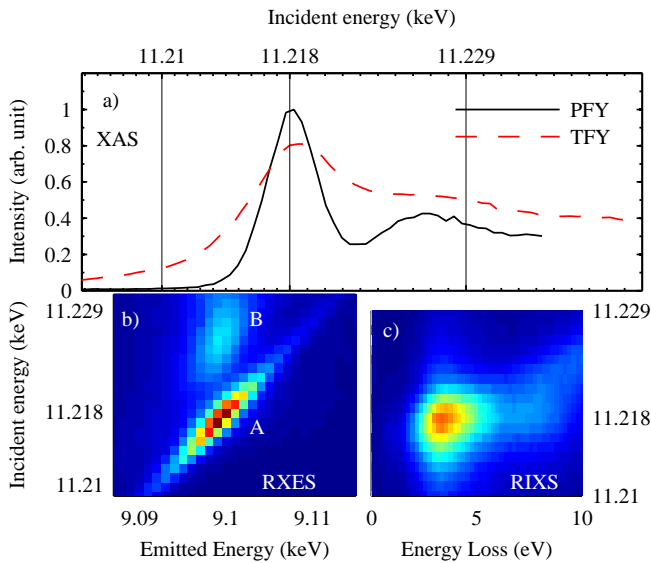}
\caption{\label{fig01}(Color online) a) Ir $L_3$-edge PFY-XAS taken by monitoring the intensity of L$\alpha_{2}$ (3d$_{3/2}$ $\rightarrow$ 2p$_{3/2}$)
as a function of incident energy. Included is the normal XAS
obtained by monitoring total fluorescence yield (TFY). b) and c)
Wide range intensity map of both RXES and RIXS spectra as a function
of the incident energy. Measurements were carried out at {\bf Q} =
(6.3 7 8.3) in the insulating phase (T = 220 K).}
\end{figure}


All x-ray measurements were carried out at the Advanced Photon
Source on the undulator beamline 9ID-B. The beam was monochromatized
by a double-crystal Si(111) and a Si(444) channel-cut secondary
crystal. For the Ir $L_3$-edge RIXS a spherical (1~m radius)
diced Si(844) analyzer was used and an overall energy resolution of
0.15~eV (FWHM) was obtained. The incident photon polarization vector
was rotated with a diamond phase plate to have it parallel to the
vertical scattering plane ($\pi$ polarization).  In this geometry, due to the polarization factor, the contribution from Thomson scattering is minimized. At this photon energy, the sample probing depth is more than 10 $\mu$m, allowing one to study true bulk electronic properties. Most of the measurements were carried out near {\bf Q} = (7 7 8) in order to
keep 2$\theta$ close to 90$^\circ$. We also utilized a
high-resolution setup with a Si(844) channel-cut secondary
monochromator and a horizontal scattering geometry. In this setup no phase plate was needed to obtain $\pi$ polarization.  This provided an
overall energy resolution of 0.08 eV (FWHM) and a significant
reduction of the elastic line. For RXES and the XAS in the partial
fluorescence yield mode (PFY-XAS) a spherical (1 m radius) diced
Ge(337) analyzer was used to provide an overall energy resolution of
0.13~eV (FWHM). No diamond phase plate was used for these
measurements. A single crystal of CuIr$_2$S$_4$ was grown by the
bismuth solution method, as previously described in
Ref.~\cite{Kiryukhin2006} and references therein. The crystal has a
triangular shape with the surface normal along the (111) direction.
Throughout this paper, we use the cubic notation ($a$ = 9.8474~$\rm{\AA}$) for simplicity \cite{Furubayashi1994}.


Wide energy range spectra obtained with all three experimental
methods are shown in Fig.~\ref{fig01}. In Fig.~\ref{fig01}(a), the XAS spectrum near the Ir $L_3$-edge is shown as a dashed line, which is obtained by
monitoring total fluorescence yield. The PFY-XAS spectra shown as
the solid line is the incident energy dependence of the L$\alpha_2$
emission line (3d$_{3/2}$ $\rightarrow$ 2p$_{3/2}$) intensity. The
benefit of PFY mode is quite clear from this figure. By suppressing
the spectral broadening due to 2p core-hole lifetime \cite{Groot2002}, sharper features are observed. The wide-range RXES
and RIXS data are presented in the bottom panels of Fig.~\ref{fig01}. In Fig.
~\ref{fig01}(b), the detailed evolution of the L$\alpha_2$ emission line as a function of incident energy ($E_i$) and emitted energy ($E_f$) is shown. In
Fig.~\ref{fig01}(c), a RIXS intensity map is plotted as a function of $E_i$ and
energy loss ($\hbar\omega \equiv E_i - E_f$).

\begin{figure}[htb]
\includegraphics[width=\columnwidth]{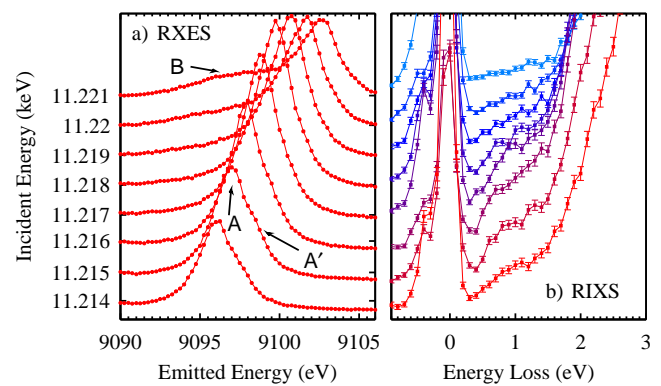}
\caption{\label{fig02}(Color online) a) Incident energy dependence of RXES spectra
taken at T = 250 K. b) Incident energy dependence RIXS spectra taken at T = 220 K and with {\bf Q} = (6.3 7 8.3). Data was taken with incident photon energy of 11.214 keV $\leq$ $E_i$ $\leq$ 11.221 keV. The spectra have been shifted to match the scale on left.}
\end{figure}

The RXES data in Fig.~\ref{fig01}(b) show two distinct features.
There is a broad, strong feature (A) with maximum intensity around
$E_i$ = 11.218~keV and another feature (B) that resonates at higher
incident energy. While feature B shows normal fluorescence behavior
with fixed energy emission, the emitted photon energy of feature A
shifts with $E_i$. This indicates that feature A comes from exciting an
Ir 2p electron into an unoccupied {\em bound} state and thus the
outgoing photon has a fixed energy loss. In this so-called Raman
regime the RXES spectra can be understood as the unoccupied density of states, resonantly enhanced over the core hole width \cite{Krisch1995}. Feature A then
corresponds to empty Ir 5d \emph{e$_{g}$} states. The evolution of
feature A as a function of $E_i$ at T = 250~K is shown in Fig.~\ref{fig02}(a). On the high energy side of A there is a prominent shoulder,
A$^\prime$, which can be associated with empty \emph{t$_{2g}$}
states. Note that the A$^\prime$ feature is resonantly enhanced for
$E_i$ that is 2-3~eV below the resonance of feature A. The intensity
ratio can be explained by the fact that there are more empty
\emph{e$_{g}$} states than \emph{t$_{2g}$} states (naively, 4
\emph{e$_{g}$} states and 0.5 \emph{t$_{2g}$} states). This
intensity ratio, in addition to the 2p core-hole broadening, is why
the \emph{t$_{2g}$}  state has not been revealed in previous XAS
studies \cite{Croft2003,Kijima2009} and in fact was not observed in
our PFY-XAS. Only by utilizing the resonance enhancement of RXES can one see this state.

\begin{figure}[htb]
\includegraphics[width=\columnwidth]{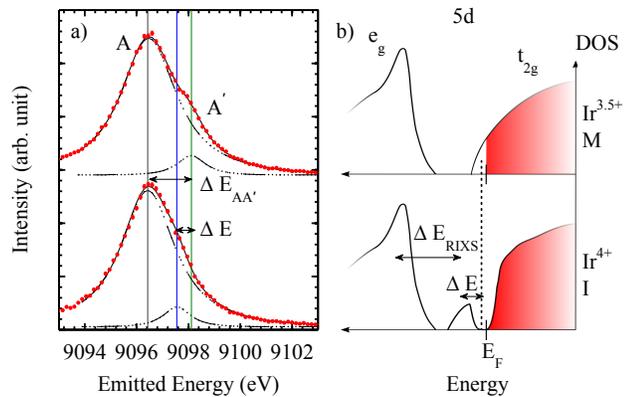}
\caption{\label{fig03}(Color online) (a)  L$\alpha_2$ RXES taken
above and below the MIT at $E_i$ = 11.215 keV. The dashed curves are
the result of a fit(see text). $\Delta E$ represents the shift of
A$^\prime$. (b) Schematic diagram of the DOS  below (I) and above (M)
the MIT inferred from our experiment. $\Delta E_{RIXS}$  represents
the splitting seen in RIXS.}
\end{figure}

Fig.~\ref{fig03}(a) shows the L$\alpha_2$ RXES spectra taken at
$E_i$ = 11.215~keV below and above the MIT. In the metallic phase,
the L$\alpha_2$ emission line clearly has two components, A and
A$^\prime$, while only A is seen in the insulating phase. The
L$\alpha_2$ emission line at T = 250~K has been fitted with two
Voigt functions, representing A and A$^\prime$. The separation of
the two peaks is estimated from the fit to be $\Delta E_{AA^\prime}
\approx$ 1.65~eV \cite{Nakai2004}. As we go below the MIT (T = 215~K) the emission line appears as a single peak A; nevertheless we could
extract A$^\prime$ using the same two-peak Voigt functions as for
the T = 250~K case. Here we have assumed that the instrumental
resolution and lifetime broadening are constant across the MIT.
The result shows almost no shift for A but the A$^\prime$ feature
shifts significantly by $\Delta E \approx 0.55$~eV. This observation
can be consistently explained by the DFT calculation result, which
is schematically shown in Fig.~\ref{fig03}(b)
\cite{Oda1995,Sarkar2009}. The main observation in our RXES study is
that the opening of an insulating gap occurs due to a shift of the empty
\emph{t$_{2g}$} band with respect to the empty \emph{e$_{g}$} band.

In Fig.~\ref{fig01}(c), a broad and strong RIXS feature in the 2-6
eV range can be seen, which shows large resonant enhancement near
$E_i$ = 11.218~keV. The weak higher energy (7-9~eV) feature is a
fluorescence line. In between these high energy excitations and the
elastic peak ($\hbar\omega$ = 0) we observe a sharp edge-like RIXS
feature around $\hbar\omega \approx$ 0.75~eV, which is clearly seen
in the individual scans shown in Fig. 2(b). This feature seems to
resonate at about 2~eV below the resonance energy of the 2-6~eV
feature. This resonance behavior is crucial in identifying the
observed spectral features. As in the interpretation of the RXES
data, the intermediate states accessed through the main absorption
feature at $E_i$ = 11.218~keV are the Ir 5d \emph{e$_{g}$} levels,
and therefore the 2-6 eV excitation can be naturally associated with a
transition into empty bands of hybridized Ir 5d \emph{e$_{g}$} and S
3p character \cite{Sarkar2009,Matsuno1997}. The 0.75~eV feature,
resonating around 11.216~keV, can then be identified as the d-d
transition into an empty \emph{t$_{2g}$}  state. The energy difference
between these two excitations represents the splitting between the empty
\emph{e$_{g}$} and \emph{t$_{2g}$} states and is found to be about
2.8 eV. This splitting is also consistent with the DFT calculation
value of about 2.6 eV \cite{Oda1995}.

\begin{figure}[htb]
\includegraphics[width=\columnwidth]{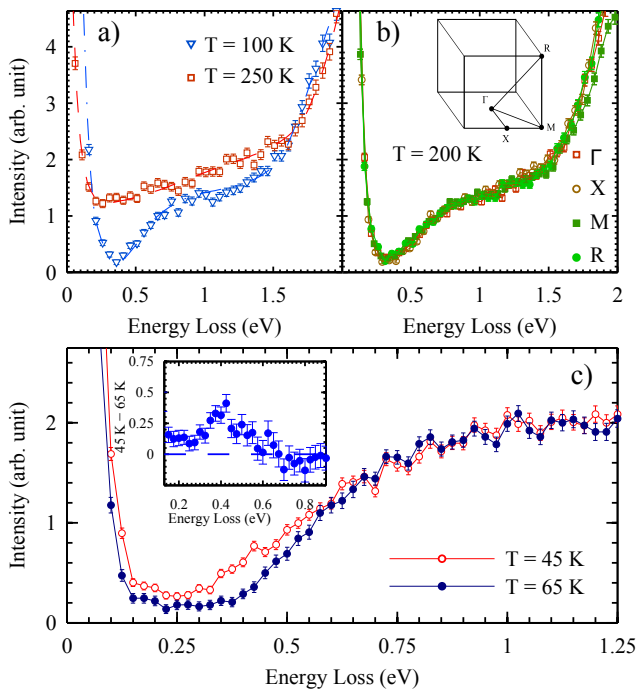}
\caption{\label{fig04}(Color online) (a) Temperature dependence of
the RIXS signal taken at {\bf Q} = (6.15 7 8.5) for $E_i$ =
11.216 keV. (b) Momentum dependence as measured at the high symmetry
positions indicated in the inset. (c) High-resolution RIXS
spectra taken above (T = 65 K) and within (T = 45 K) the charge
disordered phase. To emphasize the difference induced by irradiation
we show the subtracted spectra in inset.}
\end{figure}

In Fig.~\ref{fig04}(a) we show  in detail the temperature dependence
of the low energy RIXS spectrum.  The temperature dependence shows
drastic spectral change as the sample goes through the  MIT. In the
metallic phase at T = 250 K, a non-zero intensity is observed at low
energies. Once we go through the MIT a gap opens around 0.3 eV in the RIXS spectrum indicating the insulating state, which is consistent with the
optical study \cite{Wang2004}. According to the DFT calculation, in
the high-temperature metallic phase, hybridized Ir \emph{t$_{2g}$}
and S 3p bands cross the Fermi level \cite{Oda1995}, giving rise to
the low energy spectral weight observed below the gap. Below T = 230
K, an insulating gap opens in this band. The difference in the absolute
level of the spectral weight above the gap is probably due to the
difference in resonance condition, since electronic structure
changes quite dramatically across the MIT. The momentum dependence
of the insulating gap can be seen in Fig.~\ref{fig04}(b). It was
taken along high-symmetry directions as shown in the inset,
where $\Gamma$ is {\bf Q} = (7 7 8). Scans at the high-symmetry
positions, shown in Fig.~\ref{fig04}(b), exhibit no {\bf q}
dependence. This indicates that the gap opening is more or less
uniform in reciprocal space, suggesting that the bandwidth of
the empty \emph{t$_{2g}$} band is fairly narrow. We have also measured
the temperature and momentum dependence of the 2-6 eV feature (not
shown), but no significant changes were seen.

When the sample was cooled below T = 50~K while irradiated with
x-rays, a sudden drop of resistivity was observed
\cite{Ishibashi2002,Furubayashi2003}. It was found that not only
x-ray, but also electron, visible light, and even high energy ion
beams could induce this IILT phase
\cite{Ishibashi2002,Kiryukhin2006,Takubo2005,Koshimizu2009}. We
have obtained high-resolution RIXS data in this phase, shown in
Fig.~\ref{fig04}(c), by cooling the system with x-rays impinging on the sample. A considerable increase of the spectral weight
below 0.6-0.7~eV was observed. By subtracting the T = 65~K spectrum
from the T = 45~K one, we can see that this intensity comes from the
peak formed around 0.4~eV (shown in the inset) rather than from the tail
of the elastic peak. The spectrum does not resemble the one at T =
300~K either, which indicates that the IILT phase is different from
the original high-temperature metallic phase, in agreement with
structural studies \cite{Bozin2011}. We also measured the
L$\alpha_2$ RXES spectra for $T < 50$~K, but no difference  was
observed. We note that the larger elastic intensity ($\hbar\omega$
$\lesssim$ 0.15~eV) at T = 45~K compared to T = 65~K is due to
the increased diffuse scattering when the long range CO is
lost\cite{Kiryukhin2006}.


Even when the long-range CO is destroyed, it is known
that individual dimers do exist in this phase \cite{Bozin2011}.
One can imagine that when there is no
long-range order, dimers can hop around more easily, and the increased
hopping rate of dimers could be responsible for the decreased
resistivity in this phase. In fact such dynamic dimers could be
responsible for the mid-infrared (MIR) peak seen around 0.4 eV,
which is reminiscent of the polaron peak observed in 
manganites \cite{Kim1998}. Since dimers of Ir$^{4+}$~-~Ir$^{4+}$ have
significantly smaller bond lengths than Ir$^{3+}$~-~Ir$^{3+}$, they will introduce lattice distortion as well as charge-order defects to the crystal as they move around incoherently. Such an incoherent motion of dimers could be
strongly coupled to the lattice, resulting in formation of a
bipolaron-like object. Further high resolution spectroscopic study
of this x-ray induced phase, such as optical conductivity, could
shed light on this possibility.

In summary, we report comprehensive bulk-sensitive x-ray
spectroscopy studies of electronic structure in CuIr$_2$S$_4$. We
show that the Ir 5d \emph{t$_{2g}$} band shifts as a result of the opening of
a charge-density-wave gap, which is consistent with band theory
calculations. In the low-temperature x-ray induced phase, we found
that a broad peak around 0.4 eV appears in the RIXS spectrum, which
could arise from the dynamic hopping of local dimers.

\acknowledgements{We would like to thank Y. B. Kim, N. Perkins, and
H. Takagi for valuable discussions. Research at the University of
Toronto was supported by the NSERC of Canada, Canadian Foundation
for Innovation, and Ontario Ministry of Research and Innovation. Use
of the Advanced Photon Source was supported by the U. S. DOE, Office
of Science, Office of Basic Energy Sciences, under Contract No.
W-31-109-ENG-38.}

\end{document}